\documentclass[aps,prb,twocolumn,preprintnumbers,amsmath,amssymb,superscriptaddress,floatfix]{revtex4}
\usepackage[dvips]{graphicx}
\usepackage{mathptmx}
\usepackage{verbatim}

\begin{document}
\title{Theory of the Magnetic-Field-Induced Insulator in Neutral Graphene}
\author{J. Jung} 
\affiliation{Department of Physics, University of Texas at Austin, 78712 USA}
\author{A. H. MacDonald}
\affiliation{Department of Physics, University of Texas at Austin, 78712 USA}
\date{\today{}}

\begin{abstract}
Recent experiments have demonstrated that neutral graphene sheets have an insulating 
ground state in the presence of an external magnetic field.  We report on a
$\pi$-band tight-binding-model Hartree-Fock calculation which examines the 
competition between distinct candidate insulating ground states.  
We conclude that for graphene sheets on substrates the ground state is most likely a field-induced spin-density-wave,
and that a charge density wave state is possible for suspended samples.
Neither of these density-wave states support gapless edge excitations.
\end{abstract}

\maketitle

\section{Introduction}
The magnetic band energy quantization properties of graphene sheets lead 
to quantum Hall effects\cite{novoselov,pkim,pkim1} (QHEs) with $\sigma_{xy} = \nu e^2/h$ 
at filling factors $\nu = 4 \left( n + 1/2\right) = \cdots, -6, -2, 2, 6, \cdots $ for any 
integer value of $n$.  The factor of $4$ in this 
expression accounts for a graphene sheet's two-fold valley and spin
degeneracies.  When Zeeman spin-splitting of Landau levels is included,
quantum Hall effects are expected at the remaining even integer values of $\nu$,
including the neutral graphene $\nu=0$ case.  The $\nu=0$ quantum Hall effect of 
neutral graphene systems is interesting from two-different points of view.
First of all, the transport phenomenology of the quantum Hall effect\cite{pkim1}
is different at $\nu=0$ because of the possible absence of edge states.
Indeed the initial experimental indications\cite{pkim1} 
that a $\nu=0$ quantum Hall effect occurs in neutral graphene 
did not exhibit either the clear pleateau in $\rho_{xy}$ or the deep minimum in 
$\rho_{xx}$ which are normally characteristic of the QHE.   
Secondly, although a quantum Hall effect is expected at $\nu=0$ 
even for non-interacting electrons, the large energy gaps identified 
experimentally suggest that interactions play a substantial role in practice.
Gaps due entirely to electron-electron interactions in ordered states are 
in fact common\cite{girvin,jungwirth} in quantum Hall 
systems when two or more Landau levels are degenerate.
Partly for this reason, a number of different scenarios have 
been proposed\cite{nomura,alicea, abanin, fuchs, herbut, goerbig, yang, exciton, sheng, nomura1,sarma}
in which the gap at $\nu=0$ is associated with different types of broken symmetry
within the four quasi-degenerate Landau levels near the Fermi level of a neutral graphene sheet.  
The prevailing view has been that the ground state is spin-polarized, with  
partial filling factors $\nu_{\sigma}$ equal to $1$ and $-1$ for majority and minority 
spins respectively.  This state has an interesting edge state structure identical 
to that of quantum-spin-Hall systems,\cite{qsh} and transport properties in the quantum Hall regime that are  
controlled by the properties of current-carrying spin-resolved chiral edge-states.\cite{abanin,fb_luttinger}

The simplest picture of strong-field physics in nearly-neutral graphene sheets
is obtained by using the Dirac-equation continuum model and neglecting interaction-induced
mixing between Landau levels with different principal quantum number $n$.  
In this model, electron-electron interactions are valley and spin-dependent.
When Zeeman interactions and disorder are neglected, the broken symmetry ground state consists\cite{yang} of 
two-filled $n=0$ Landau levels with arbitrary spinors in the $4$-dimensional spin-valley space.
This family of states is favored by electron-electron interactions because of Fermi 
statistics which lowers Coulomb interaction energies when the orbital content of electrons in the 
fermion sea is polarized.  When Zeeman coupling is included, it uniquely
selects from this family the state in which both $n=0$ valley orbitals are occupied for majority-spin
states and empty for minority-spin states.  The interacting system ground state is then 
identical to the non-interacting system ground state, although interactions are 
expected\cite{yang} to dramatically increase the energy gap for charged excitations. 

This argument for the character of the ground state appears to be compelling, but its conclusions
are nevertheless uncertain.  First of all, Landau-level mixing interactions are normally 
stronger than Zeeman interactions, and could play a role \cite{fertigLLmix}.
In addition, although corrections\cite{alicea,goerbig,abanin_2} to the 
continuum model for graphene are known to be small at experimental field strengths, they 
could still be more important than the Zeeman interactions.
Suspicions that the character of the ground state could be misrepresented by
the $n=0$ continuum model theory have been heightened recently by the work 
of Ong and collaborators, who found a steep increase in the Dirac point resistance \cite{ong} 
with magnetic field and evidence for a field-induced transition to a strongly insulating 
state at a finite magnetic field strength.\cite{ong1}
Somewhat less dramatic increases in resistance at the Dirac point have 
also been reported by other researchers.\cite{exp1,exp2,exp3}

In this article we attempt to shed light on the ground state of neutral graphene  
in a magnetic field by performing self-consistent Hartree-Fock calculations 
for a $\pi$-orbital tight-binding model.  In the continuum model, Hartree-Fock
theory is known\cite{yang} to yield the correct ground state.  By 
using a $\pi$-orbital tight-binding model we can at the same time 
conveniently account for Landau-level mixing effects and systematically
account for lattice corrections to the Dirac-equation continuum model. 
As we will discuss at length below, it is essential to perform the 
mean-field-theory calculations with Coulombic electron-electron interactions, and not 
the Hubbard-like interactions commonly
used \cite{herbut,tchougreeff} with lattice models.
One disadvantage of our approach is that our calculations are feasible only
at magnetic fields strengths which are stronger than those available experimentally.
We therefore carefully examine the dependence on magnetic field
strength and extrapolate to weaker fields. 
We conclude that  under typical experimental conditions the most-likely 
field-induced state of neutral graphene on a SiO$_2$ substrate 
is an spin-density-wave state,
and that suspended samples might have a charge density wave state. 
Neither of these orderings support edge states in the $\nu=0$ gap.
We also discuss the magnetic field dependence of different contributions to the total energy
and estimate a critical value of perpendicular and tilted magnetic field at which Zeeman splitting 
will bring about a phase transition to a solution with net spin polarization which 
{\em does} support edge states.

Although our calculation
captures some realistic features of graphene sheets that are 
neglected in continuum models, it is still not a 
complete {\em all electron} many-body theory.  In particular we neglect the 
carbon $\sigma$ and $\sigma^*$ orbitals whose polarization is expected to screen
the Coulomb interactions at short distances.  Because of our uncertainty as to the strength of 
this screening, our conclusions cannot be definitive.  We nevertheless hope
that our calculations, in combination with experiment, will prove useful
in identifying the character of the field-induced insulating state in neutral graphene.

Our paper is organized as follows.  In Section ~\ref{model} we explain in detail the model which 
we study which has two parameters, a relative dielectric constant $\epsilon_r$ which
accounts for the dielectric environment of the graphene sheet, and on on-site interaction 
$U$ which accounts for short-distance screening effects, for example by $\sigma$-band 
polarization.  Our main results on the competition between different ordered states are
presented in Section ~\ref{results}.  In Section ~\ref{ribbons} we turn to a 
discussion of the electronic structure of neutral graphene ribbons, paying particular 
attention to their edge states which play a key role in most quantum Hall transport experiments.
Finally in Section ~\ref{discussion} we summarize our main conclusions.

\section{Interacting-Electron Lattice Model for Graphene Sheets} 
\label{model} 

\subsection{Non-Interacting Electron $\pi$-band model} 

We first comment briefly on the $\pi$-band tight-binding model of 
graphene in the presence of a magnetic field.\cite{mfield,kohmoto_graphene,momtambaux,jinwu, arikawa}   
Each carbon atom on graphene's honeycomb lattice has three near-neighbors 
with $\pi$-orbital hopping parameter $t = -2.6 eV$.  Magnetic field effects are captured by  
a phase factor in the hopping amplitudes: $t \rightarrow t \times e^{2 \pi i \phi}$
where $\phi = (e/ch) \int {\bf A} d{\bf l}$
depends on line integral of the vector potential ${\bf A}$ along a trajectory linking the two 
lattice sites.  When the dimensionless magnetic flux 
density is $\phi \equiv B \, S_{hex}  / \phi_0 = 1/q$, where $q$ is an integer and 
$\phi_0 = ch/e$ is a magnetic flux quantum,
it is possible to apply Bloch's theorem in a unit cell which is enlarged 
by a factor of $q$ relative to the honeycomb lattice unit cell.
(The honeycomb lattice unit cell area $S_{hex} = \sqrt{3} a^2 /2$ and $a = 2.46 \AA$ for a graphene sheet.)
Lattice model Landau levels have a small width 
which increases with magnetic field strength and reflects magnetic breakdown 
effects neglected in the continuum model.

The ground state energy density differences discussed below scale approximately as powers of the 
magnetic length $\ell_{B}$, defined by $2 \pi \ell_{B}^2 B = \phi_0$.  ($\ell_{B}$ and $q$ are  
related by $\ell_{B}  = \left( S_{hex} q / 2 \pi \right)^{1/2}
= 0.371 \sqrt{q} \, a =  0.913 \, \sqrt{q}  \,\,\,\, \AA  $.) 
In a continuum model description the density contributed by a single 
full Landau level is $1/2\pi \ell_{B}^2$ and the energy of the $n^{th}$ Landau level is 
given by $E_n = \pm 2 \hbar v_{F} \sqrt{|n|} / \ell_B$ 
where $v_{F} = \sqrt{3} a t/2\hbar$ is the Fermi velocity of graphene.
All energy levels evolve with magnetic field except for 
the $n=0$ level, $E_0=0$.\cite{castroneto}
When the $n^{th}$ Landau level is full it contributes 
$E_n / \left( 2 \pi \ell^2_B \right)$ to the energy density. 
From this we immediately see that in the weak-field limit 
important energies tend to scale as $\ell_{B}^{-3} \propto B^{3/2}$.
It is easy to show, for example, that the magnetic-field dependence of the 
total band energy of a neutral non-interacting graphene sheet  
is given by $E\left( \ell_B \right) = a_{kin} / \ell_{B}^{3}$
where $ a_{kin} = 2.65 \,\, eV \cdot \AA^3$. 
This non-analytic field-dependence is responsible for the divergent weak-field 
diamagnetic response ($ \left( \partial E_{tot} / \partial B \right) / B $)
of graphene discussed some time ago by McClure\cite{mcclure}.
We show below that when interactions are included, the 
energy differences between competing field-induced-insulator states also 
tend to vary as $\ell_{B}^{-3}$. 

\subsection{$\pi$-band Model Effective Interactions}

It is clear from previous analysis of lattice-corrections to 
continuum models\cite{alicea, goerbig,abanin_2,exciton} and from lattice-model
calculations based on extended Hubbard models\cite{herbut} 
that conclusions on the nature of the field-induced insulating ground state 
are very dependent on the effective electron-electron interactions 
used in a $\pi$-band lattice model of graphene.  
In particular, it seems clear that the long-range $1/r$ Coulomb interaction
tail is essential.  We approximate the interaction between $\pi$-orbitals located at sites separated by 
a distance $d$ by $V \left(  d  \right) = 1 /  ( \epsilon_{r} \sqrt{ a_o^2 + d^2} )$ 
where $a_o = a/\left(2 \sqrt{3}\right)$, the bonding radius of the carbon atoms, 
accounts approximately for interaction reduction due to $\pi$-charge smearing on 
each lattice site, \cite{interactions} and $\epsilon_{r}$ accounts for screening due 
to the dielectric environment of the graphene sheet.  (Here energies are in Hartree ($e^2/a_{B}$) units and 
lengths are in units of the Bohr radius $a_{B}$.) 
The onsite repulsive interaction parameter, $U$, is not well known and we take it to be
a separate parameter.  We motivate the range of values considered for this 
interaction parameter below. 
The value chosen for $\epsilon_r$ can also represent in part screening 
by $\sigma$ orbitals neglected in our approximation, or be understood as an {\em ad-hoc} correction for
overestimates of exchange interactions in Hartree-Fock theory.  Although we study a 
range of values for this interaction parameter model in order to test the robustness of 
our conclusions, we believe that a value of $\epsilon_r \sim 4$ is normally appropriate for 
graphene sheets placed on a dielectric substrate. 
For practical reasons we truncate the range of Coulomb interaction in real space at  
$d = L_{max} = 6.5 a$. 
This type of truncation is especially helpful when treating systems without periodic boundary conditions
and allows us to avoid problems due to slowly converging sums in real space
that can otherwise be treated through the Ewald sum method.\cite{ewald}
Truncation of the Coulomb interaction at a 
reasonably large $L_{max}$ must however be applied with utmost care in order to obtain
solutions consistent\cite{cutoff,monolayer} with the limit $L_{max} \rightarrow \infty$.

In considering appropriate values for the on-site interaction $U$ we can start from the 
Coulomb interaction energy at the carbon radius length scale which is $\sim 20 {\rm eV}$,
while this estimate can be reduced if one considers a charge distribution corresponding to a $p$-orbital.
In fact an estimate from the first ionization energy and electron affinity gives $U = 9.6 {\rm eV}$. \cite{dutta}
It is known that the effective on-site interaction strength is greatly reduced from this bare 
value in the solid state environment because of screening by polarization of bound orbitals on 
nearby carbon atoms.  We consider values of $U$ between $2 {\rm eV}$ and $6 {\rm eV}$,
bracketing values deemed appropriate by a variety of different researchers
\cite{alicea,yazyev,bhowmick,wunsch}. 
A larger value of $U$ increases the interaction energy cost of any charge-density-wave (CDW) 
state which might occur.  The direct interaction energy is zero when all carbon sites stay neutral, but 
can be positive or negative in CDW states.  In the CDW states we discuss below electron density 
$\delta n$ is transferred between A and B honeycomb sublattices.  In this state the direct interaction
energy per site is  
\begin{equation} 
\label{eq:di} 
\delta E_{DI} = \frac{(\delta n)^2}{2} \; \Big[ U + \sum_{j \in A}   
 V(d_{ij}) - \sum_{ j \in B} V(d_{i j}) \Big],
\end{equation} 
where $d_{ij}$ is the distance between lattice sites $i$ and $j$, $U = V(d_{ii})$ and $i$ is a fixed label 
belonging to sublattice $A$.
The largest terms in Eq.(~\ref{eq:di}) are the repulsive on-site interactions which are 
proportional in our model to $U$ and attractive excitonic 
interaction between electrons on neighboring opposite sublattice sites which are inversely proportional to $\epsilon_r$.  
Using an Ewald technique to sum over distant sites we find that $\delta E_{DI}$ is positive  
for $\epsilon_r \cdot U > 13.05$eV.
(The corresponding criterion for the truncated Coulomb interactions we use 
in our self-consitent-field calculations is $\epsilon_r \cdot U >  = 12.23$eV; the difference
between the right-hand-side of these two equations is one indicator of the 
inaccuracy introduced by truncating the Coulomb interaction.) 
When 
$\epsilon_r \cdot U < 12.23$eV the CDW state is stable unless 
band and exchange energies support a uniform density state \cite{monolayer}. 

Given the band structure model and the interaction model, the Hartree-Fock 
mean-field-theory calculations for bulk graphene sheets with periodic boundary conditions
and for graphene ribbons reported in the following sections are completely 
standard.\cite{szabo}  The band quasiparticles are determined by diagonalizing 
a single-particle Hamiltonian which includes direct and exchange interaction terms. 
The direct and exchange potentials are expressed in terms of the occupied quasiparticle
states and must be determined self-consistently.  (We do not quote the detailed expressions
for these terms here.)  Since the Hartree-Fock equations can be derived by minimizing the 
total energy for single Slater determinant wavefunctions, every solution we find corresponds
to an extremum of energy.  The iteration procedure is stable only if the extremum is a 
minimum so we can be certain that all the solutions found below represent local energy 
minima among single Slater determinant wavefunctions with the same symmetry properties.

\section{Field-Induced Insulating Ground States}
\label{results}

\begin{figure*}[htbp]
\begin{center}
\includegraphics[width=9cm,angle=90]{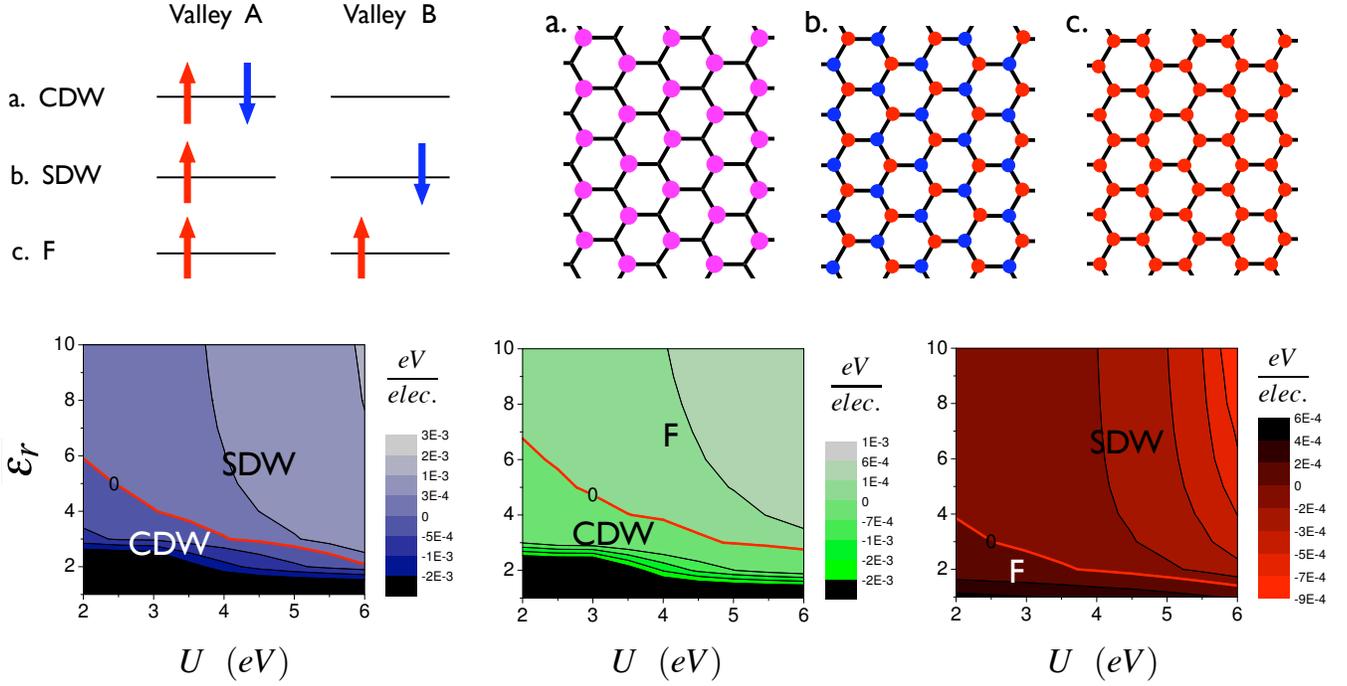} 
\caption{
(Color online)
{\em Upper panel:} Schematic representation of valley polarized charge density wave (CDW),
spin density wave (SDW) and ferromagnetic (F) spin polarized broken symmetry solutions 
that can be obtained in a self-consistent mean field calculation of graphene 
under a perpendicular magnetic field at half filling.
Each arrow represents the filling of one $n=0$ Landau level of a given
spin and valley.
{\em Lower panel:}
From left to right $E^{CDW} - E^{SDW}$, $E^{CDW}-E^{F}$ and $E^{SDW} - E^{F}$ 
total energy differences per electron in $eV$ 
as a function of the onsite repulsion $U$ and $\varepsilon_r$ obtained from a 
data mesh of 9 $\times$ 10 points, calculated neglecting the Zeeman term and 
for a magnetic field of 792 Teslas corresponding to $1/100$ of a flux quantum 
per honeycomb hexagon.
For smaller values of $U$ the $CDW$ solutions are energetically 
favored whereas for larger values of $U$ the $SDW$ solutions are 
favored in a wide range of $\varepsilon_r$. 
The $F$ solutions are never the lowest in energy.
The red contour lines indicate degeneracy between two different solutions.
 } 
\end{center}
\end{figure*}

\subsection{Identification of Candidate States} 

At zero-field band energy favors neutral graphene states  
without broken symmetries, and there is no compelling evidence from experiment that they are induced by interactions. 
In a perpendicular magnetic field, however, the systems is 
particularly susceptible to the formation of broken symmetry ground states because of the 
presence of a half-filled set of four-fold spin (neglecting Zeeman) and valley degenerate Landau levels with (essentially) perfectly quenched band energy. 
Although the final ground state selection probably rests on considerations 
that it fails to capture, the $n=0$ continuum model captures the largest part of the interaction energy 
and most of the qualitative physics.    
The ground state is formed by occupying two of the four $n=0$ 
Landau levels, selected at random from the four-dimensional orbital space,
and producing a gap for charged excitations.
 
Three representative broken symmetry states are illustrated in Fig.~1.
Because $n=0$ Landau levels orbitals associated with 
different valleys are completely localized on different honeycomb 
sublattices, 
a charge density wave (CDW) solution results when $n=0$ orbitals are occupied for both 
spins of one valley.
(When Landau level mixing is neglected valley indices 
and A or B sublattice indices are equivalent.)
This state lowers the translational symmetry of the 
honeycomb lattice in a way which removes inversion symmetry.
The other extreme is a spontaneously spin-polarized
uniform density state (ferro - F)
in which $n=0$ orbitals are occupied in both valleys but only for one spin-component.
A third type of broken symmetry state, the spin-density-wave (SDW) state, 
has both broken inversion symmetry and 
broken spin-rotational invariance.  In the cartoon version of Fig.~1, 
$n=0$ electrons occupy states with one spin-orientation on
one sublattice and the opposite spin-orientation on the other sublattice. 
Possible broken symmetry states, some at other filling factors, 
had been discussed previously by several authors.\cite{alicea,herbut}
These three states are all 
contained within the $n=0$ continuum model family of ground states whose degeneracy is lifted 
by by lattice nd Landau level mixing effects.
In the self-consistent mean-field-theory calculations described in detail 
below, the three states identified above all appear as energy extrema
in our collinear-spin study.

\subsection{Energy Comparisons} 
\begin{figure*}
\begin{center}
\begin{tabular}{cc}
\includegraphics[width=8.45cm,angle=0]{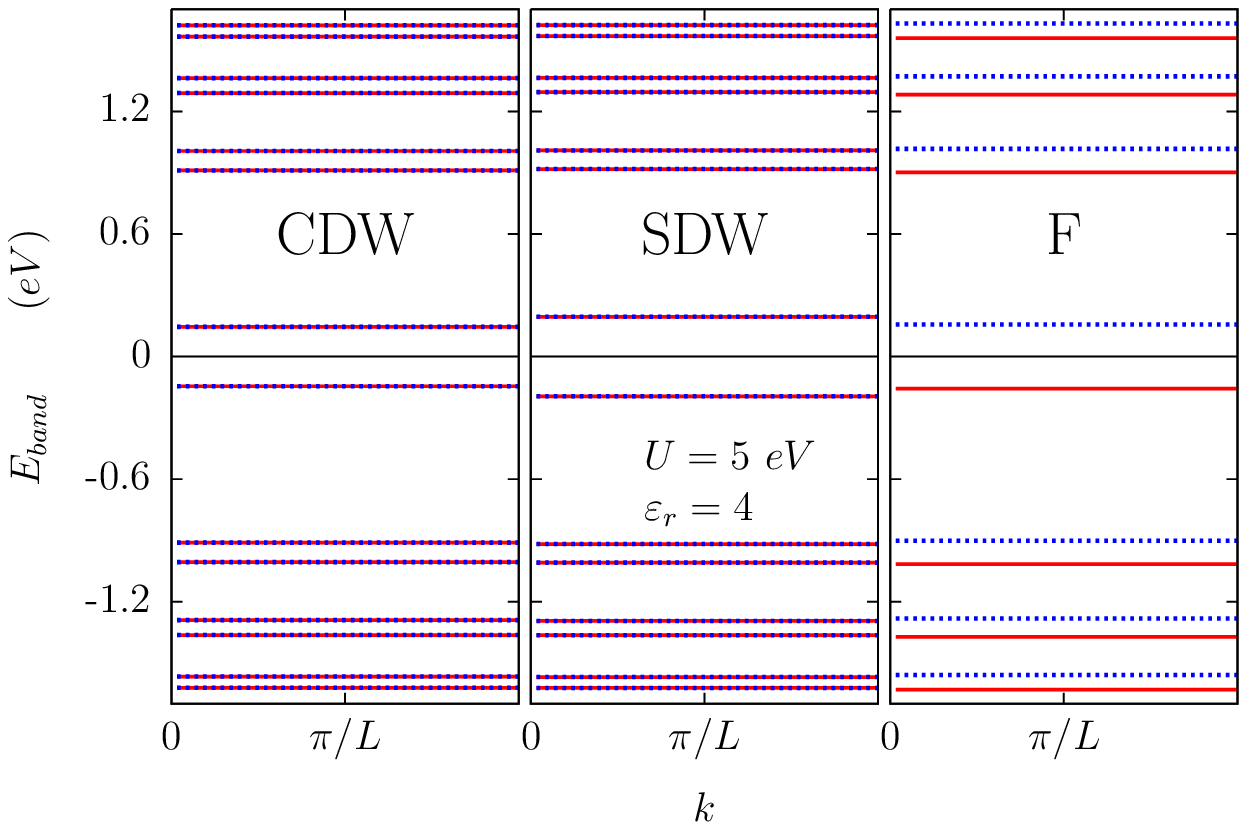}  \quad \quad \quad&
\includegraphics[width=8.15cm,angle=0]{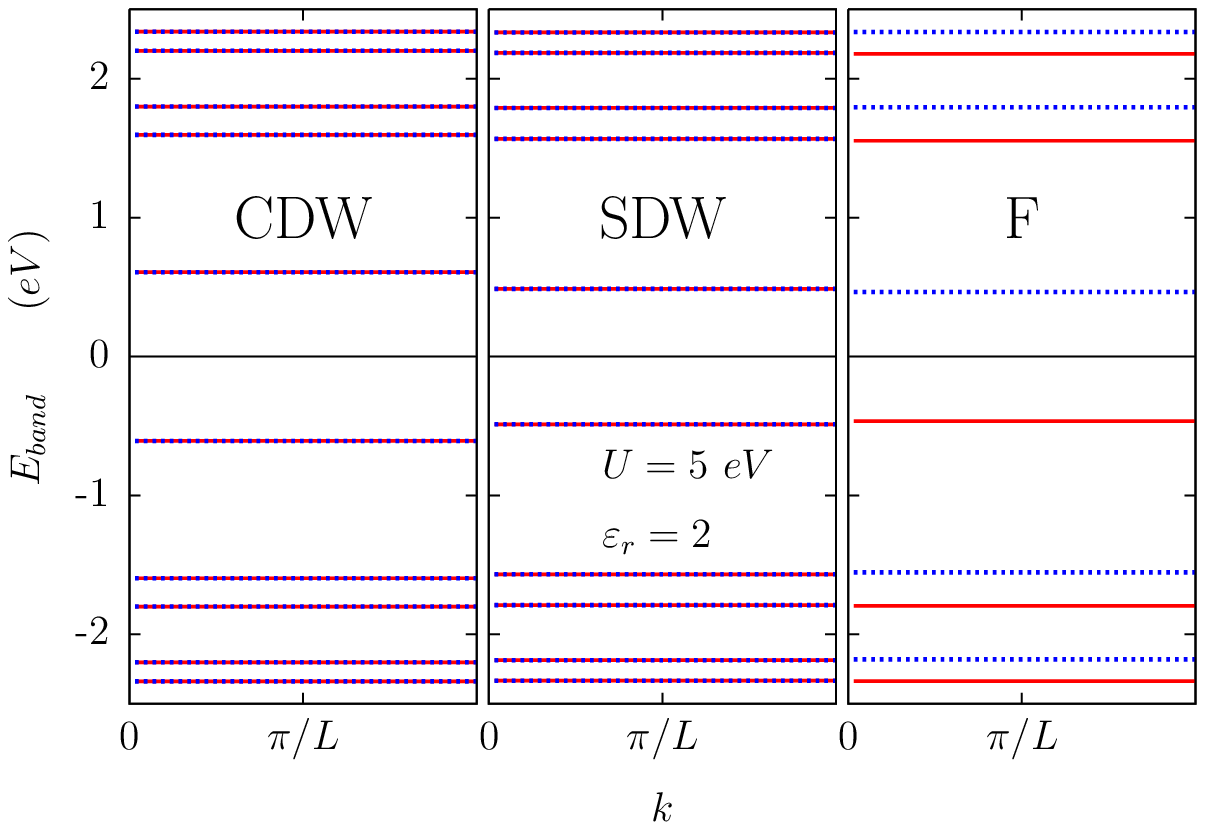} 
\end{tabular}
\end{center}
\caption{ 
(Color online) One dimensional representation of the dispersionless 
band structure of a graphene sheet
under a strong magnetic field $B = 440 T$ 
represented in the momentum coordinate $k$ parallel to the narrower direction of the 
unit cell. The band-gaps follow the $B^{1/2}$ scaling law expected from the 
continuum model.
The red color is used to represent up spin while blue is used for down spin. 
{\em Left Panel:} Band structure for CDW, SDW and F solutions obtained for  
$U = 5 eV$ and $\varepsilon_r = 4$. 
For these interaction parameters the SDW state has the lowest energy
and the largest gap at the Fermi level.
{\em Right Panel:} Band structures for $U = 5 eV$ and $\varepsilon_r = 2$. 
When the 
on site repulsion is sufficiently weak the  energetically favored 
solution corresponds to the CDW state and this solution then has the largest
gap.
}
\end{figure*}

In order to examine the physics behind the competition between the candidate 
ground states we decompose the total energy for all three contributions into 
band, direct interaction, and exchange interaction contributions.  We have obtained 
self-consistent solutions for all three states over a range of on-site interaction $U$ and the dielectric screening $\epsilon_r$ 
values. 
Because of kinetic-energy quenching in the (essentially degenerate) $n=0$ Landau level,
the interaction strengths required to drive the system into an ordered state are essentially zero.
The key question, then, is which state is favored.   
In Fig. (~1) we illustrate how the energy differences between the 
three states depend on the model interaction parameters.  
The results in this figure were obtained for $q = 100$ unit cells per flux quantum,
which corresponds to perpendicular field strength $B = 792$ Tesla. 
The unit cells in which we can apply periodic boundary conditions in this 
case contain $100 \times 2$ lattice sites.  The $k$-space 
integrations in the self-consistent Hartree-Fock calculations were performed using
a $60$ $k$-point Brillouin zone sampling.  The self-consistent field equations
were iterated until the total energies were converged to nine significant figures.
High accuracy is required because the three states are very similar in energy since
the ordering occurs primarily
in the $n=0$ Landau level, involving only $1\%$ or so of the electrons for this value of $q$.
This accuracy was sufficient to evaluate energy differences that typically have
three significant figures. 

The first point to notice in these plots of energy differences is that the 
two uniform charge density solutions, the F solution and the SDW solution, 
behave similarly.  The largest contrast therefore is between the CDW 
solution and the SDW and F solutions.  Focusing first on the CDW/SDW comparison 
we notice that the SDW state is favored when $U$ is large or $\epsilon_r$ is large.
The crossover occurs near $\epsilon_r \cdot U \sim 12 \,\, eV$,  very close to the 
line along which $\delta E_{DI}$ changes sign.  The fact that the CDW/SDW phase boundary occurs very 
close to this line is expected because of kinetic energy quenching in a magnetic field.    
When the non-uniform density CDW state is compared with the uniform-density spin-polarized 
F state the phase boundary moves very close to a larger value of this product with 
$\epsilon_r  \cdot U $ ranging from $\sim 14$ to $\sim 18 \, {\rm eV}$ along the phase boundary.
Evidently the competition between CDW and SDW states is based very closely on the direct interaction energy,
with additional weaker elements of the competition entering when the F state is considered.  

Direct comparison between the uniform density SDW and F solutions indicates that the 
latter is favored only at values of $U$ and $\epsilon_r$ which are outside the 
range of most likely values.  
As discussed in more detail below, we find that the direct interaction energy in these 
two states is identical, and that the more negative exchange 
energy of the SDW state overcomes a larger band energy.  
In this case the main difference between the energies of the two states arises
from Landau level mixing effects.  As we explain later,  
Landau level mixing leads to a local spin-polarization 
which is larger in the SDW state than in the F state.

In Fig. (~1) we have introduced the main trends in the energetic competition between 
CDW, SDW, and F states.  However, as we have explained, these calculations were undertaken
at field strengths that exceed those available experimentally.  In the following subsection 
we demonstrate that the field dependence of the energy comparisons is extremely systematic so that 
extrapolations down to physical field strengths are reliable.  So far we have also ignored Zeeman
coupling which favors F states.  This coupling can be important and is also addressed in the 
following subsection.

\subsection{Field Strength and Zeeman Coupling Dependence} 

\begin{figure}
\begin{center}
\begin{tabular}{cc}
\includegraphics[width=4cm]{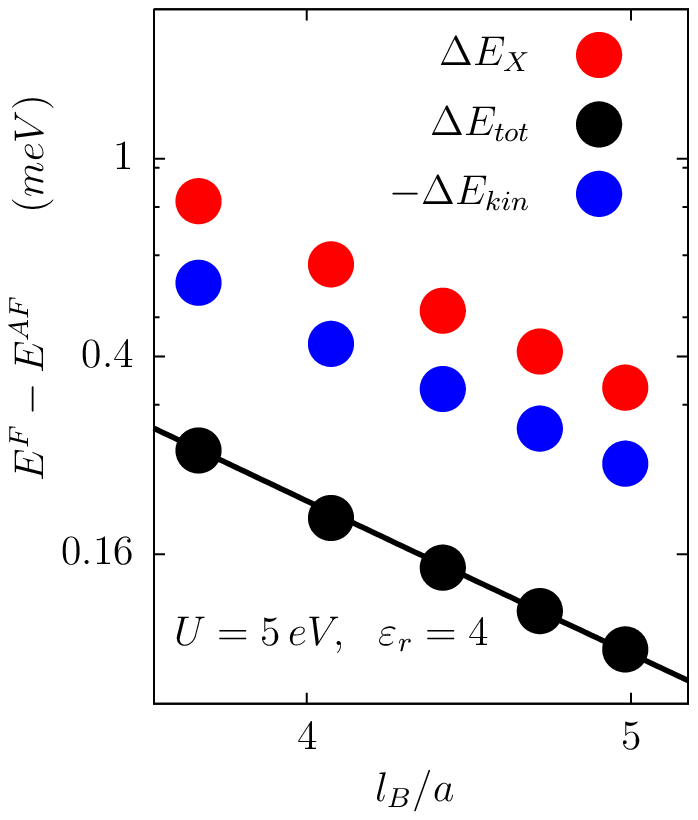} \quad \quad &
\includegraphics[width=4cm]{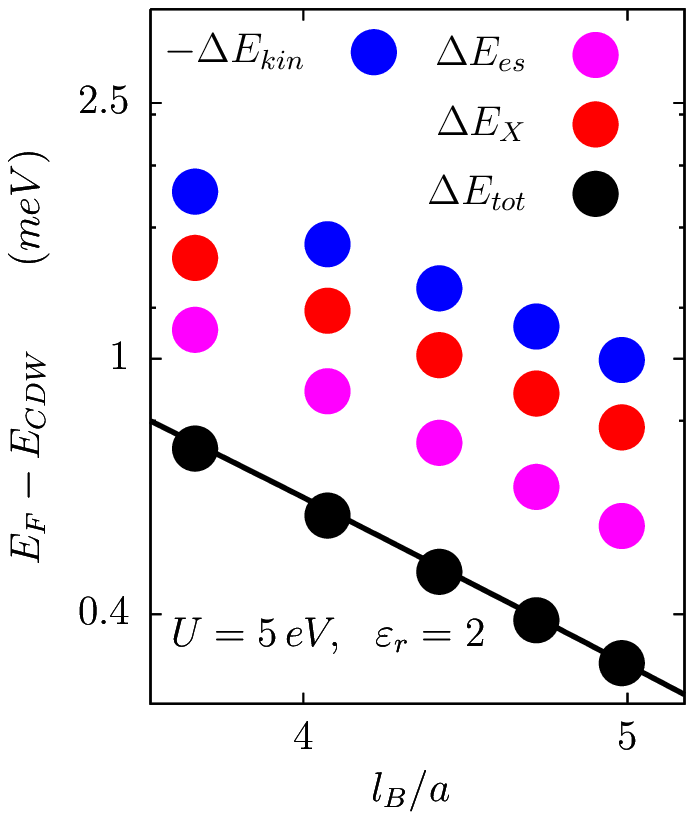} \\ 
\end{tabular}
\end{center}
\caption{(Color online)
Total energy per site differences $\Delta E_{tot}$, 
separated into kinetic energy $\Delta E_{kin}$,
electrostatic energy $\Delta E_{DI}$ and exchange energy $\Delta E_{XI}$ contributions
as a function of magnetic length $\ell_{B}$ in lattice constant $a = 2.46 \AA$ units.  
The total energy differences were fitted
to a $C_{3/2} B^{3/2} + C_{2} B^{2}$ curve.  The fitting parameters are listed in Table I.
{\em Left panel:}
Energy differences between F and SDW solutions $\Delta E^{F/SDW} = E^{F} - E^{SDW}$.
These results were obtained with interaction 
parameter values $U = 5 \,\, eV$ and $\epsilon_r = 4$ for which SDW is the lowest energy configuration.
The more negative values of exchange energy in the SDW state compensates the kinetic 
energy penalty related to the inhomogeneous accumulation of the electron wave functions at
alternating lattice sites. The electrostatic energy differences are zero thanks to 
the uniform electron density for both solutions.
{\em Right panel:}
Same as the previous figure but for $\Delta E^{F / CDW} = E^{F} - E^{CDW }$.
The interaction parameters in this case are $U = 5 \,\, eV$ and $\epsilon_r = 2$ for which 
CDW is the lowest energy configuration. 
When the onsite repulsion $U$ is small enough that the electrostatic energy penalty
for the inhomogeneous charge distribution is small, exchange is the main 
contribution driving the CDW instability. However, in the case illustrated here 
$U$ is so small that the electrostatic part of the hamiltonian does play an important role
in favoring the CDW state.
The energy contributions follow a magnetic field decay law that 
deviates more from $B^{3/2}$ than in the previous case 
because the on-site interaction $U$ plays an essential role. 
}
\end{figure}
\begin{figure}[htbp]
\begin{center}
\includegraphics[width=8cm]{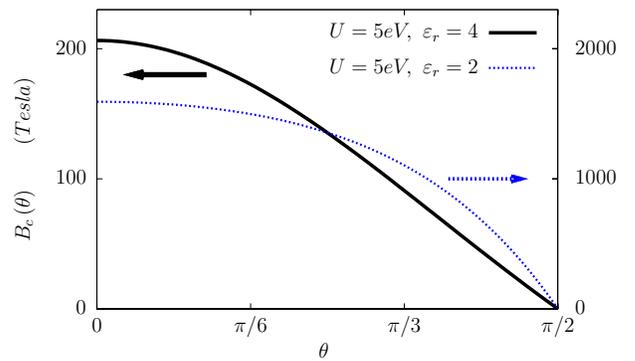} 
\caption{
Tilt angle $\theta$ dependence of the critical magnetic field required to induce a transition to the F state.
The black solid curve and dashed blue curve represent the critical magnetic fields starting from the SDW and CDW states respectively.
In the CDW curve we observe a larger deviation from a simple $\cos \left( \theta \right)$ law
due to a stronger influence of lattice scale physics described by the $C_2$ coefficient in equation 
\ref{bcritical}.} 
\end{center}
\end{figure}
\begin{table*}[ht] 
\centering 
\begin{ruledtabular}
\begin{tabular}{|c | c | c | c| c | c| c | c| c | c |c | } 
\multicolumn{1}{c|}  {}               & 
\multicolumn{2}{c|} {$\epsilon_r = 2 $}       &
\multicolumn{4}{c|} {$\epsilon_r = 3 $}       &
\multicolumn{4}{c}  {$\epsilon_r = 4$ }        \\ 
\hline                    
\hline 
\multicolumn{1}{c|} {U}               & 
\multicolumn{1}{c|} {$C^{CDW}_{3/2}$ }       &
\multicolumn{1}{c|} {$C^{CDW}_{2}$}       &
\multicolumn{1}{c|} {$C^{CDW}_{3/2}$ }       &
\multicolumn{1}{c|} {$C^{CDW}_{2}$ }       &
\multicolumn{1}{c|} {$C^{SDW}_{3/2}$ }       &
\multicolumn{1}{c|} {$C^{SDW}_{2}$ }       &
\multicolumn{1}{c|} {$C^{CDW}_{3/2}$ }       &
\multicolumn{1}{c|} {$C^{CDW}_{2}$ }       &
\multicolumn{1}{c|} {$C^{SDW}_{3/2}$ }       &
\multicolumn{1}{c|} {$C^{SDW}_{2}$ }        \\  \hline 
2 & NA  & NA               &  322    &   -1.89   &   ---      &  ---         &  52.6     &   1.05      & ---         & ---  \\
3 & NA  & NA               &  130    &    0        &   ---      &  ---         &  23.2     &   0.0314  & ---         & ---  \\
4 & 1070    & -13.6      &  54.2   &  -0.210    &   ---     &  ---        &   ---       &   ---          &   51.6    &  0.0209    \\    
5 & 417    & -3.15        &  ---        &  ---         &   111   &  -0.630   &  ---       &  ---           &  92.8    &  0.839       \\ 
6 & 198    & -1.67        &  ---        &  ---         &   255  &  -2.10     &  ---       &   ---          &   241     & 0           
\\ [0.5ex]  
\hline
\hline
\multicolumn{1}{|c|} { $U$ }               & 
\multicolumn{2}{c|} {$B_C^{CDW } $}      &
\multicolumn{2}{c|} {$B_C^{CDW } $}      &
\multicolumn{2}{c|} {$B_C^{SDW } $}      &
\multicolumn{2}{c|} {$B_C^{CDW } $}      &
\multicolumn{2}{c|} {$B_C^{SDW } $}        \\  \hline 
\multicolumn{1}{|c|} { 2 }               & 
\multicolumn{2}{c|} {NA}      &
\multicolumn{2}{c|} {1200}      &     
\multicolumn{2}{c|} {---}      &
\multicolumn{2}{c|} {70}      &  
\multicolumn{2}{c|} {---}        \\  
\multicolumn{1}{|c|} { 3 }          & 
\multicolumn{2}{c|} {NA}      &
\multicolumn{2}{c|} {300}      &   
\multicolumn{2}{c|} {---}      &
\multicolumn{2}{c|} {10}      &   
\multicolumn{2}{c|} {---}        \\  
\multicolumn{1}{|c|} { 4 }               & 
\multicolumn{2}{c|} {2600}      & 
\multicolumn{2}{c|} {52}      &   
\multicolumn{2}{c|} {---}      &
\multicolumn{2}{c|} {---}      &
\multicolumn{2}{c|} {50}        \\    
\multicolumn{1}{|c|} { 5 }               & 
\multicolumn{2}{c|} {1600}      & 
\multicolumn{2}{c|} {---}      &
\multicolumn{2}{c|} {200}      & 
\multicolumn{2}{c|} {---}      &
\multicolumn{2}{c|} {200}        \\    
\multicolumn{1}{|c|} { 6 }               & 
\multicolumn{2}{c|} {490}      &  
\multicolumn{2}{c|} {---}      &
\multicolumn{2}{c|} {740}      &  
\multicolumn{2}{c|} {---}      &
\multicolumn{2}{c|} {1100}      
\\ [0.5ex] 
\end{tabular} 
\end{ruledtabular}
\label{table:nonlin} 
\caption{ 
In the upper table we list the values of $C_{3/2}^{\left( SDW/ CDW \right)}$ in units of  $10^{-10} eV/T^{3/2}$
and $C_{2}^{\left( SDW/ CDW \right)}$ in units of $10^{-10} eV/T^{2}$,
obtained from fits to the their energy difference with respect to F states as given in equation (\ref{tiltedfield})
for two different values of the interaction parameters $U$ and $\varepsilon_r$. 
The critical field $B_c$ estimates are dependent on the reliability of these fits.
Note that a crossover between SDW and CDW states can be driven by changes in the 
dielectric screening environment captured by $\varepsilon_r$.
The lower table lists $B_c$ values at which the F states become critical according to 
Eq.(\ref{bcritical}).
 } 
\end{table*} 

We now turn our attention to the magnetic field dependence of the solutions.
For this purpose we found self-consistent solutions over a range 
of magnetic fields for two sets of interaction parameters, 
$U = 5 eV$ and $\varepsilon_r = 4$  for which the SDW solution
has the lowest energy,
and $U = 5 eV$ and $\varepsilon_r =2$ for which the CDW solution
has the lowest energy.
The band structures of the different possible solutions for these set of parameters
are shown in Fig. (2) and the field dependences of the three energy differences
are plotted in Fig. (3).  We 
see that every contribution accurately follows a $B^{3/2} \propto l_{B}^{-3}$ law
with small deviations that can be accounted for by allowing a term proportional to $B^{2}$. 
This is the same field-dependence law that we discussed earlier for the 
case of a non-interacting electron system.  In the continuum model it
is guaranteed in neutral graphene when electrons interact via the  
Coulomb interactions by the fact that both kinetic and interaction energy 
densities then scale as $(length)^{-3}$; the magnetic field simply provides a 
scale for measuring density.  The fact that we find this field dependence
simply shows that the condensation energies of all three ordered states
are driven by continuum model physics. 
This is in agreement with the intuitive picture of the interaction energy as the product of
the number of electrons occupying a Landau level which is 
directly proportional to $B$, multiplied by the Coulomb interaction scale 
for electrons in the $n=0$ Landau level which is proportional to $B^{1/2}$.
The fact that the differences in energy between the three states 
follows this rule suggests that the most important source of 
differences in energy between these states is Landau-level mixing, which should not 
violate the $B^{3/2}$ law.
Small deviations from this law are expected because of lattice effects.
The deviations are stronger in CDW solutions than in the SDW solutions because
of the charge density inhomogeneity at the lattice scale present in the former.

We can draw two important additional conclusions from the
$B^{3/2}$ behavior.  First of all, lattice effects
are not dominant effect at the field strengths
for which we are able to perform calculations, and should be less
important at the weaker fields for which experiments are performed 
because the magnetic length $l_{B}$ will then be even longer compared
to the honeycomb lattice constant.  The difference in energy between
the three states should mainly vary as $B^{3/2}$ all the way down to 
zero field, provided only that disorder is negligible.  (We discuss
the role of disorder again in Sec.~\ref{discussion}.)  Our calculations
should therefore reliably predict the energetic ordering of the states 
in the experimental field range.  The second conclusion we can make 
concerns the importance of Zeeman coupling which we have ignored to
this point. 
First of all, Zeeman coupling will have a negligible effect on the energies of the SDW and 
CDW states since they have a vanishing spin magnetic susceptibility.  The energy 
difference per site between the F state and the two 
density-wave states can be written in the form 
\begin{eqnarray}
\Delta E &=& E^{\left(SDW/CDW \right)}  - E^{F}  \nonumber \\
&=&  B^{3/2} C_{3/2}^{\left( SDW/CDW \right)}  \cos(\theta)^{3/2}  \nonumber  \\
&+&  B^2 \cdot \left(  C^{\left( SDW/CDW \right)}_{2} \cos \left( \theta \right)^2 -  C_{Z} \cos(\theta)  \right) 
\label{tiltedfield} 
\end{eqnarray}
where $B$ is the total magnetic field strength and $\theta$ is the field 
tilt angle relative to the graphene plane normal.  Factors of $B \cos(\theta)$ in this 
expression therefore account for the perpendicular field dependence.  
The second term in Eq.~\ref{tiltedfield} contain the contributions that scale with $B^{2}$.
The factor of $B \cos(\theta)$ which appears in the Zeeman term is present because the spin-polarization
of the F state is proportional to the Landau level degeneracy.  
The coefficients $C_{3/2}$ and $C_{2}$ can be obtained by fitting energy differences obtained 
from numerical solutions of the self-consistent field equations, like those plotted in 
Fig.~3, and depend on the interaction model parameters as shown in Table I.  
The Zeeman coefficient in Eq.~\ref{tiltedfield} is $C_{Z} = 7.3 \cdot 10^{-10} \,\, eV / T^2$ is 
independent of interaction parameters.  From the above equation we find that the $F$ state 
has lower energy than the spin-unpolarized states for 
\begin{equation}
\label{bcritical}
B >   B_{c} \left( \theta \right) = \frac{C_{3/2}^2  \cos(\theta)  }{ \left( C_{z}  - C_2 \cos (\theta) \right)^2}  
\end{equation} 
The fields required to achieve an energetic preference for the spin-polarized
state are smaller at larger tilt angles because the 
orbital energy has a stronger $\theta$ dependence.

In table I we show the values of $C_{3/2}$ and $C_{2}$ for SDW and CDW configurations
favored with respect to F for a set of parameters of $U$ and $\epsilon_r$.
We notice that the coefficients dictating the critical field transition to F solutions
can be made relatively small if the parameters are near the 
crossover boundary to F states.
It is possible that a SDW or CDW to F transition could be induced 
by varying magnetic field.  If a transition was observed, most likely by a 
change in transport properties as discussed in the next section, 
it could provide valuable input on the effective interaction parameters of the 
$\pi$-orbital tight-binding model.

\section{Quantum hall edge states in graphene ribbons}

\begin{figure}[htbp]
\begin{center}
\includegraphics[width=6.5cm,angle=90]{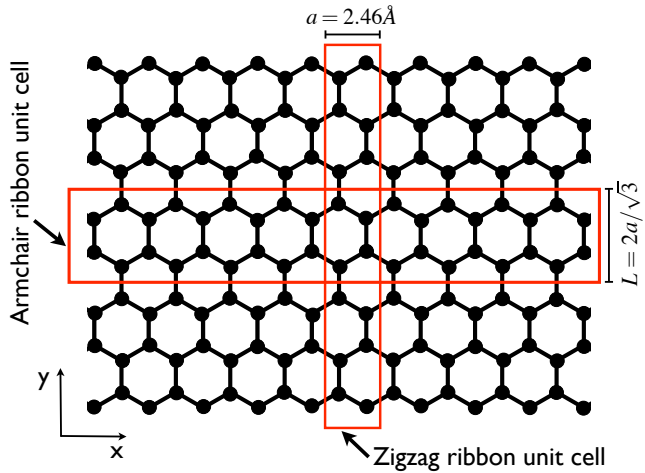} 
\caption{
Representation of the unit cell choices for armchair and zigzag edge terminated graphene nanoribbons.
} 
\end{center}
\end{figure}

\begin{figure*}[t]
\label{ribbons} 
\begin{center}
\begin{tabular}{ccc}
\includegraphics[width=6.1cm,angle=0]{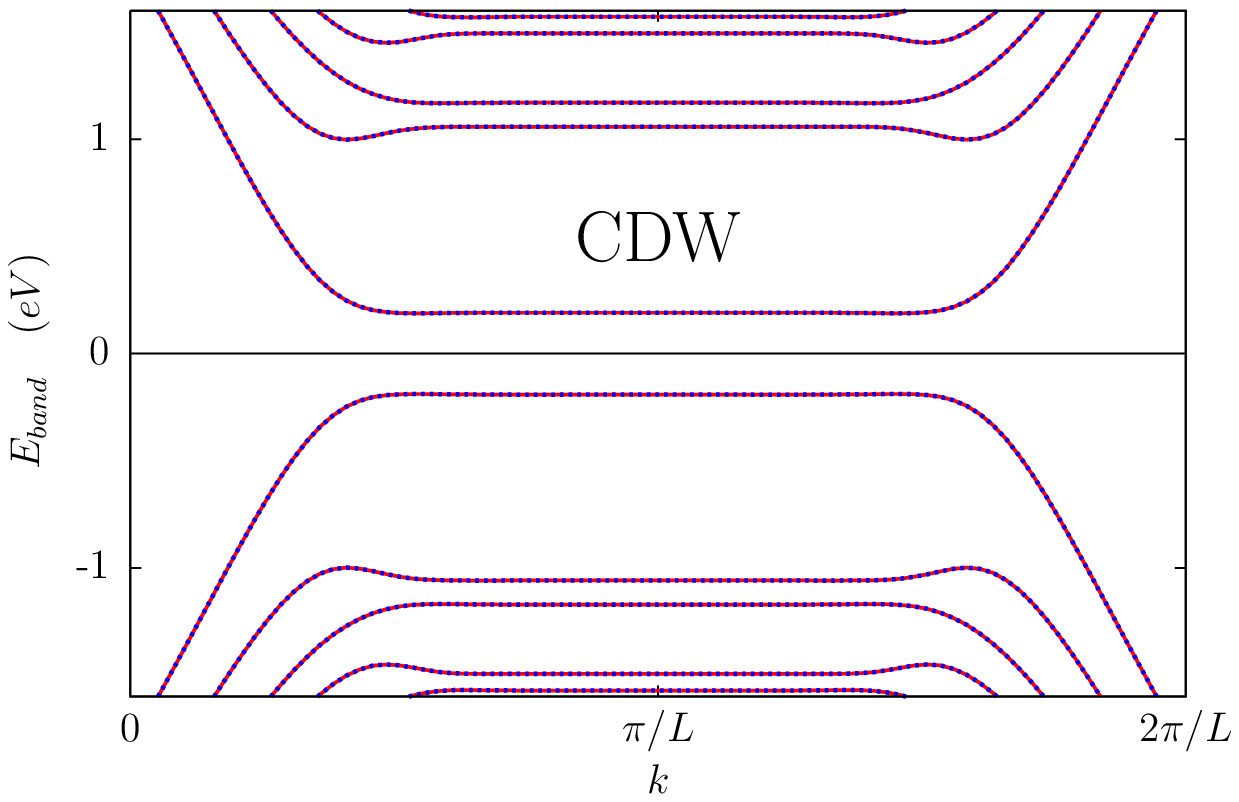}  $\,\,\,\,$ &
\includegraphics[width=5.7cm,angle=0]{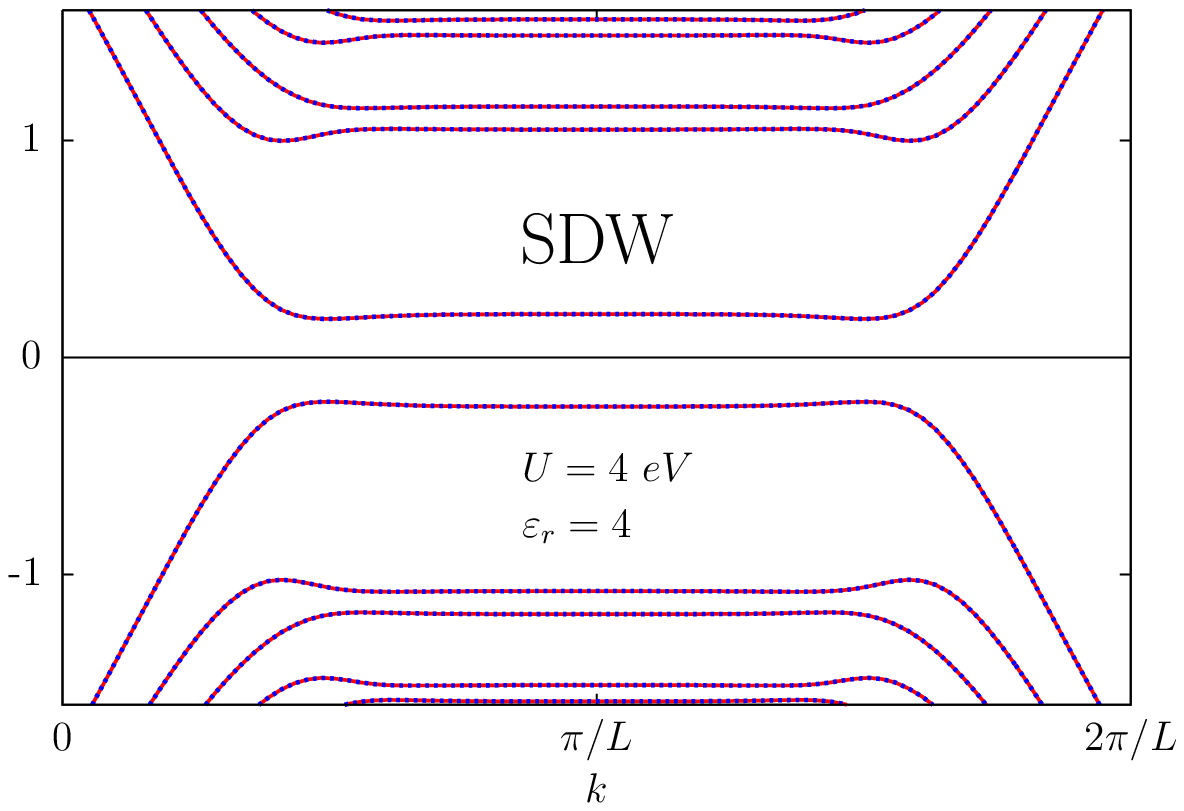}  &
\includegraphics[width=5.7cm,angle=0]{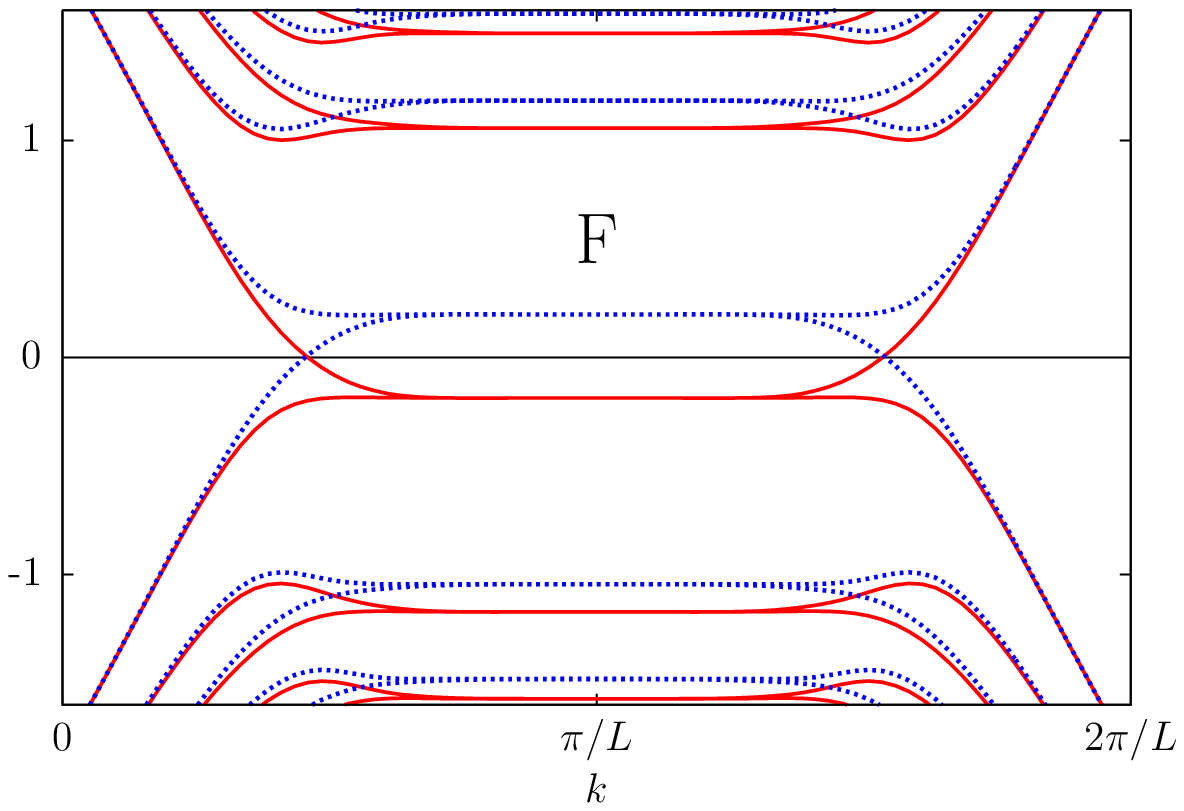}  \\ \\
\includegraphics[width=5.9cm,angle=0]{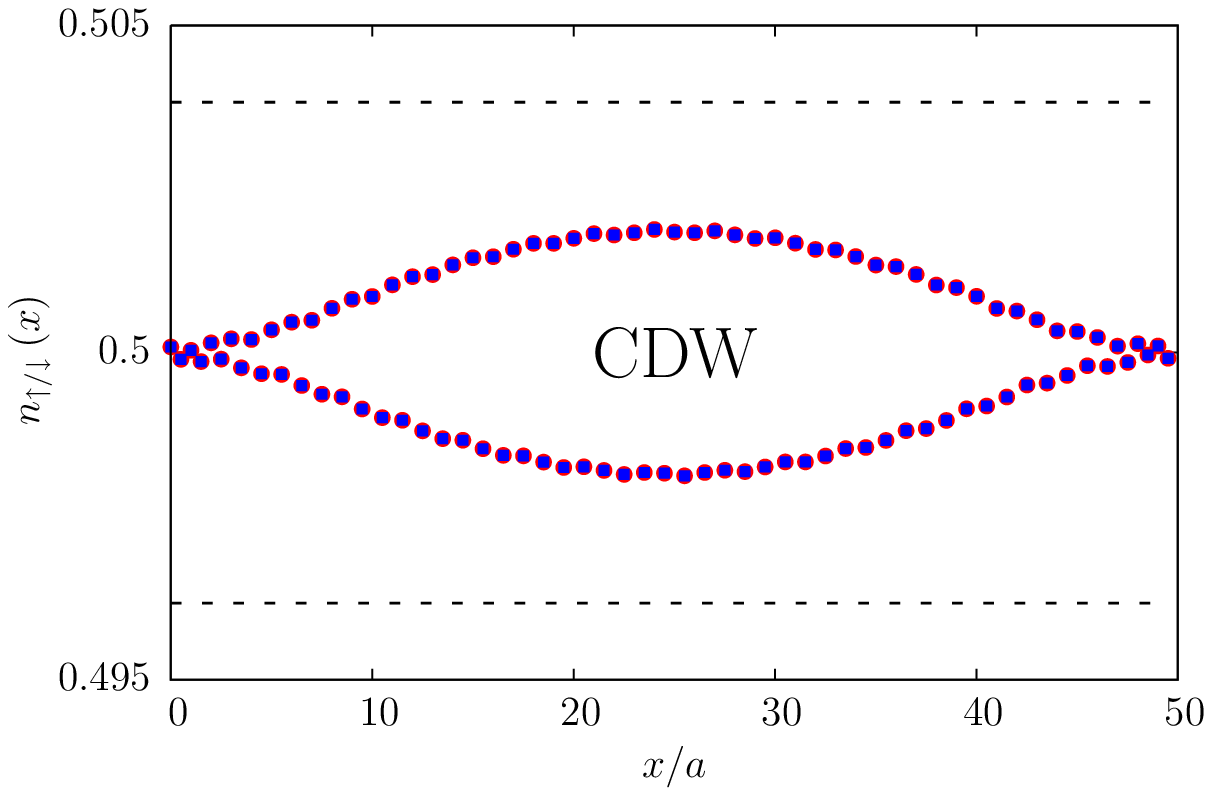} $\,\,\,\,$ &
\includegraphics[width=5.5cm,angle=0]{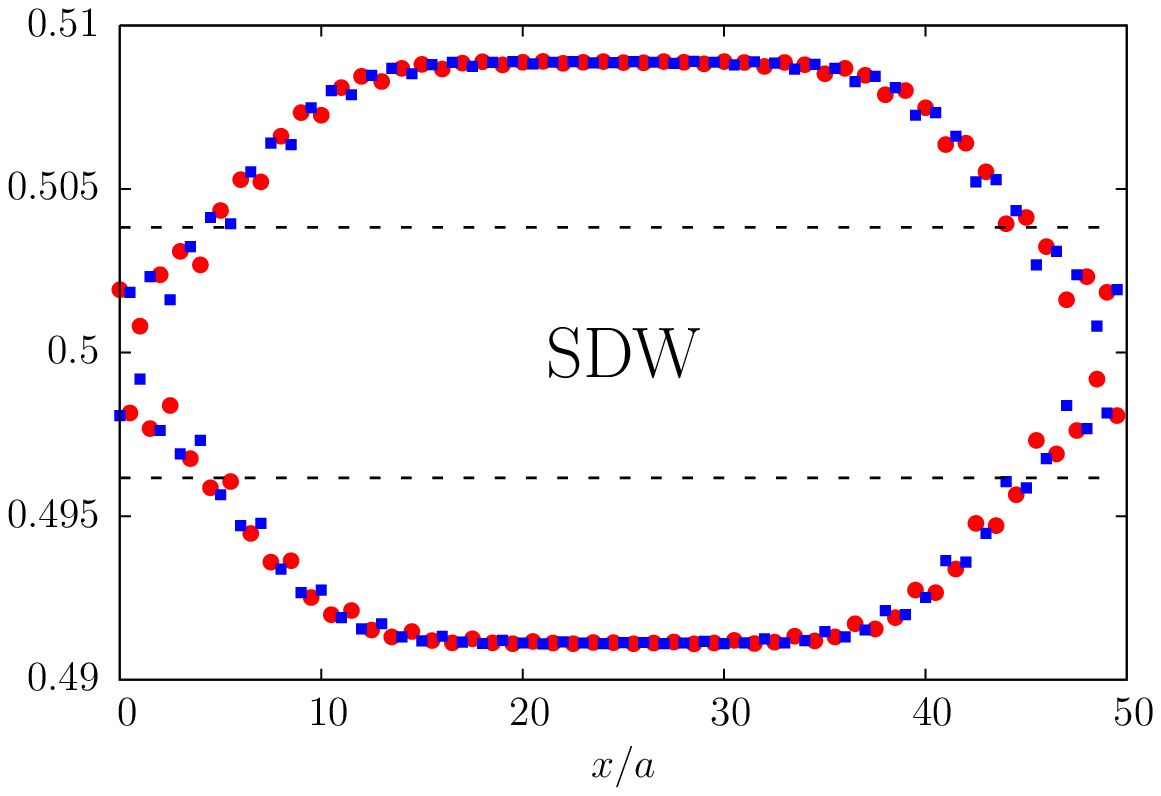}  $\,\,$  &
\includegraphics[width=5.5cm,angle=0]{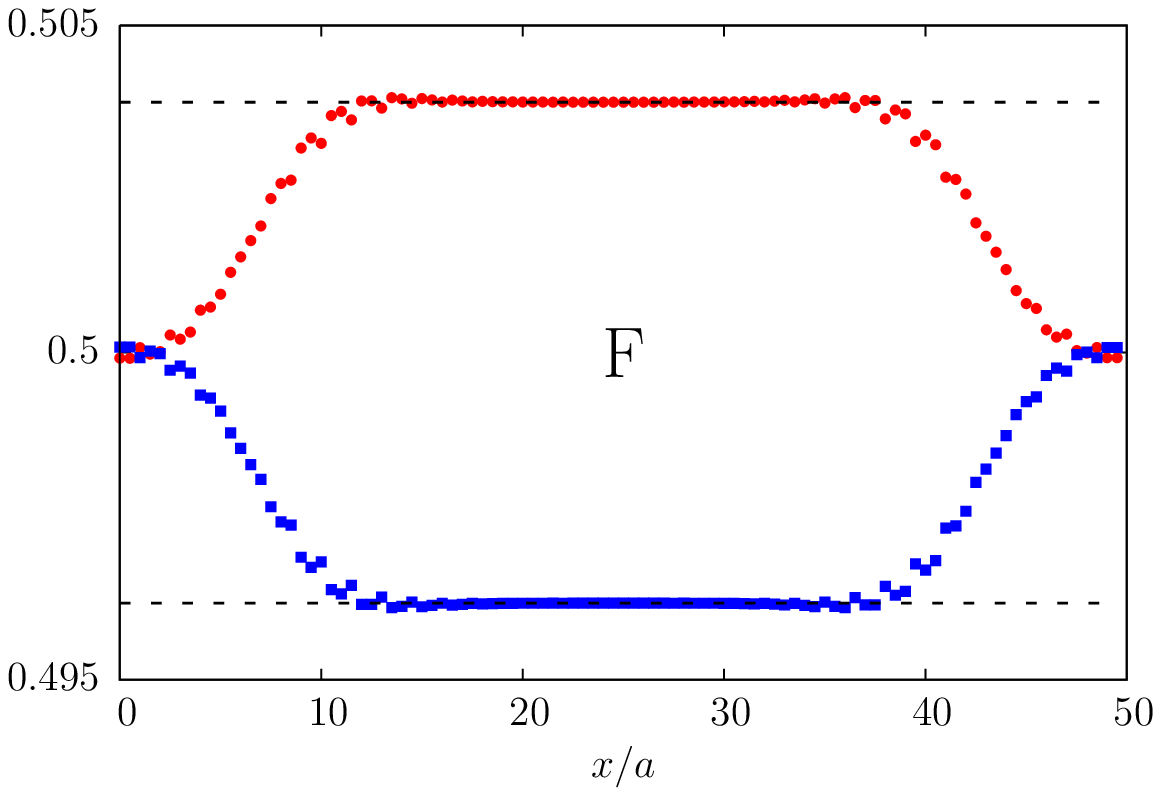}   $\quad$
\end{tabular}
\end{center}
\caption{ 
(Color online) Band structure and spin density distributions in armchair ribbons calculated 
for $U=4\,\, eV$ and $\epsilon_r = 4$ for a magnetic field corresponding to $\ell_B = 4.24 a$ (
B = 606 Tesla) 
using a $100 \times 2$ lattices in the unit cell and $80$ k-point sampling in the Brillouin zone. 
The red color is used to represent up spin while blue is used for down spin. 
{\em Upper row:} Armchair ribbon band structures for the  F, SDW and CDW solutions.
Only the F solution is metallic, while the other two configurations are insulating. 
For this choice of parameters SDW is the energetically favored state, 
but the energies of the other states are similar and their band gaps also have a similar size.
{\em Lower row:} Spin-$\uparrow$ and spin-$\downarrow$ density distributions across one of the zigzag rows 
in the unit cell of an armchair ribbon.
The influence of Landau-level mixing on the three solutions
is apparent in the differences between the three sets of density distributions. 
In the case of AF solutions Landau-level mixing enhances the local spin-density,
while in the case of the CDW solution the magnitude of the charge density oscillation 
is reduced by Landau level mixing.
The dashed horizontal lines represent the occupation in absence of Landau level mixing.}
\end{figure*}

The quantum Hall effect occurs when a two-dimensional electron system has a chemical potential 
discontinuity (a gap for charged excitations) at a density which depends on magnetic field.
A gap at a field-dependent density necessarily\cite{halperin,macdonaldleshouches} implies the presence of chiral edge states
that support an equilibrium circulating current.  The current varies with chemical potential 
at a rate defined by the field-dependence of the bulk gap density.
Most quantum Hall measurements simply reflect the property\cite{macdonaldleshouches} that separate local equilibria are 
established at opposite edges of a ribbon in systems with a bulk energy gap or mobility gap.   
It is immediately clear therefore that 
the $\nu=0$ quantum Hall effect is special since it is due to an energy gap at the neutrality point,
{\em i.e.} at a density which does not depend on magnetic field.  The issue of whether or not 
the $\nu=0$ gap and associated phenomena should be referred to as an instance of the quantum 
Hall effect is perhaps a delicate one.  The $\nu=0$ gap is intimately related to Landau quantization 
and in this sense is comfortably grouped with quantum Hall phenomena.  This view supports the 
language we use in referring to the $\nu = 0$ quantum Hall effect.  On the other hand,
since it occurs at a field-independent density, its transport phenomena are 
more naturally viewed as those of an ordinary insulator\cite{sarma} which just happens to be 
induced by an external magnetic field.

An exception occurs for the F state which does have edge states\cite{abanin,fb_luttinger},
and can be viewed as having $\nu=1$ for majority spins and $\nu=-1$ for minority spins. 
In the simplest case, it has two branches of edge state with opposite chirality for 
opposite spin, much like those of quantum-spin-Hall\cite{qsh} systems.  
In a Hall bar geometry most transport measurements are very strongly sensitive
to the presence or absence of edge states.  
In order to address edge state physics directly at $\nu=0$ 
we have extended our study from the bulk graphene to the graphene nanoribbon case.
Tight-binding model solutions for a ribbon in the presence of a magnetic field 
can be obtained in essentially the same way as for bulk graphene, 
with the simplification that any magnetic field strength preserves the one-dimensional ribbon wavevector 
$k$ as a good quantum number when the gauge is chosen appropriately.
This graphene ribbon problem in a magnetic field was studied time ago by Wakabayashi {\em et. al.} 
\cite{waka_bfield} and revisited recently within both tight-binding \cite{arikawa, huang}
and continuum \cite{abanin,brey_bfield,gusynin_edge} models in order to provide 
a microscopic assessment of the relationship between Landau levels and edge states.
The general feature of the ribbon band structure in the presence of a magnetic field
is that those states localized near the edges have dispersive bands,
whereas those in the flat band region are located mostly in the bulk. 
In the case of zigzag edge termination, 
edge localized states are present even in the absence of a magnetic field \cite{fujita}.
In the quantum Hall regime these states are in the non-dispersive band region like the bulk localized states
and they do not contribute to 
edge currents, although they can interact with other edge localized states.

Fig.~6 explicitly illustrates how the character of the bulk broken symmetry 
is manifested in ribbon edge state properties. 
Because of practical limitations our calculations are restricted to  
moderately narrow ribbons with widths of order $10 {\rm nm}$. 
In order to properly reproduce bulk Landau 
level quantization in these narrow systems we have to choose 
magnetic field strengths strong enough to yield magnetic lengths
$\ell_{B} \sim 25 {\rm nm}/(B[{\rm Tesla}])^{1/2}$ substantially smaller than the 
ribbon width, {\em i.e.} fields stronger than typical experimental fields.
On the other hand if the magnetic fields are too strong, say 
$\ell_B < a$    
the levels will be strongly affected by the lattice and the properties of the solutions 
will substantially depart from the behavior we should expect at weaker magnetic fields,
for which the continuum model description is approximately correct.
For field strengths in the appropriate range, we find the same three 
types of self-consistent field solutions as in the bulk calculations, namely
F, CDW, and SDW solutions. 
The band structures and spin resolved densities presented in Fig.~6 confirm that 
only the F solution has states in the broken-symmetry induced gap.
Hall-bar transport properties for the F configuration have been 
discussed by Abanin {\em et al.} \cite{abanin} and Fertig {\em et al.}\cite{fb_luttinger}
from a theoretical point of view.   
The CDW and SDW state electronic structure is insulating, both at the edge and 
in the bulk.  The band structures of these two-states are similar even though 
their spin-density profiles are quite distinct.

The plots of spin-$\uparrow$ and spin-$\downarrow$ partial densities across the ribbons 
hint at some of the physics which selects between the three candidate ordered states.
In the truncated $n=0$ Landau-level continuum theory, the F state has one excess 
occupied Landau level for majority spins and one deficiency in occupied Landau levels
for minority spins.  We see in Fig.~6, that the size of these polarizations is 
not strongly influenced by the Landau level mixing effects included in our 
lattice calculation.  The $n=0$ continuum theory SDW state has the same spin excesses
and deficiencies, but they have opposite sign on opposite sublattices.  We see
in Fig.~6, that these order parameters are actually enhanced by Landau level mixing 
effects; inter-Landau level exchange effects polarize lower-energy occupied 
Landau level states so that they enhance the SDW pattern.  For the CDW state on the other hand,
the $n=0$ Landau level excess density on one sublattice is suppressed by Landau-level mixing.
In this case the direct electrostatic interaction is non-zero so that the 
occupied Landau levels away from the Fermi energy are polarized between
sublattices in the opposite sense of the $n=0$ levels.  The fact that the SDW 
state is enhanced by Landau level mixing explains why it is favored over the 
F state for the interaction parameters used to construct Fig.~6.
 
Finally, we comment on the microscopic electronic structure at the edges 
of an F state.  As illustrated in Fig. 6, 
the mean-field  
electronic structure at the edge contains a domain wall.\cite{fb_luttinger}
Because textures in this domain wall can\cite{fb_luttinger} carry charge, the 
F state edge is however not necessarily insulating in the absence of disorder.
It was recently argued that impurities with magnetic moments near the  
sample edges can introduce spin-flip backscattering potentials \cite{fb_luttinger} 
whose effectiveness is enhanced by the presence of a domain wall.
The high resistances seen experimentally at $\nu=0$ therefore do not necessarily 
prove that the ground state is not an $F$ state.  
The analysis in Ref.\onlinecite{fb_luttinger} was carried out within a 
Luttinger liquid formalism whose parameters depend on the domain wall shape.  
As illustrated in Fig. 7, 
our microscopic domain walls exhibit an interesting 
anisotropy in which inner and outer segments of the wall differ.

\begin{figure}
\label{hybridized}
\begin{center}
\begin{tabular}{cc}
\includegraphics[width=4.7cm]{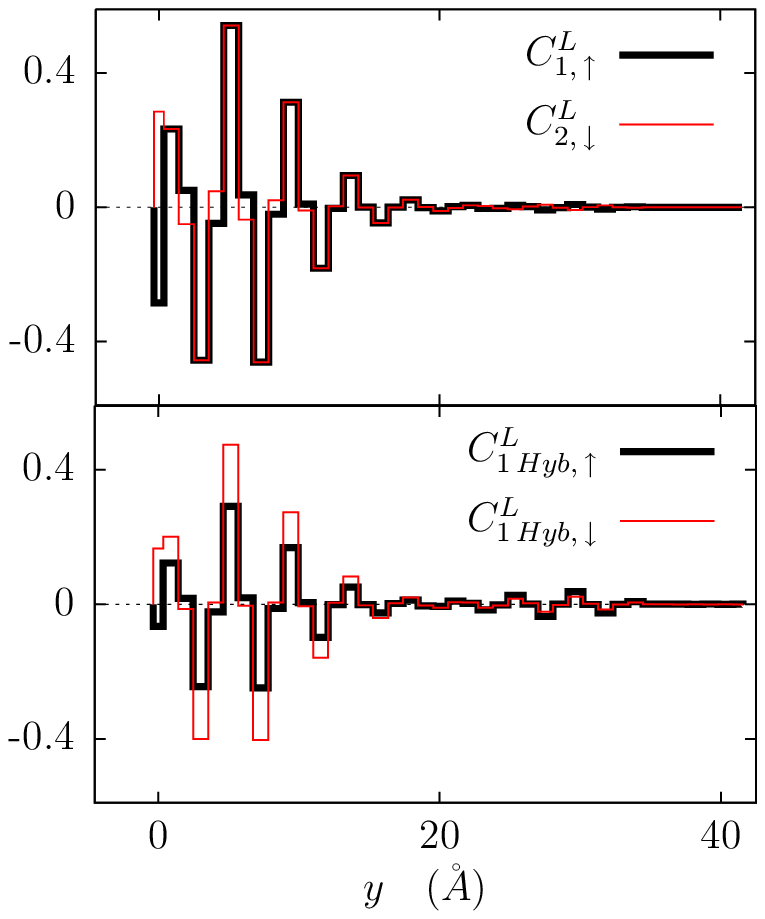}  \,\, &
\includegraphics[width=3.5cm]{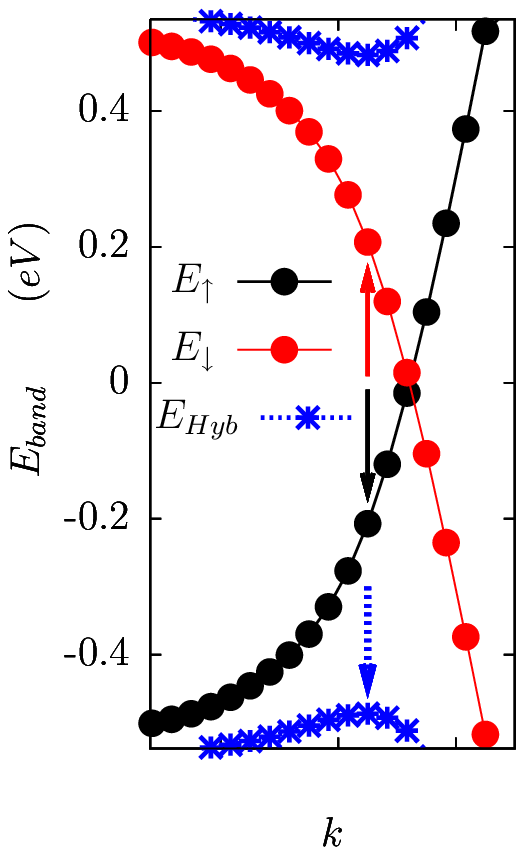}   
\end{tabular}
\end{center}
\caption{ 
(Color online)
Quantum hall edge domain-wall wave functions and energy bands
corresponding to an $F$ solution
calculated for a zigzag graphene nanoribbon with 20 carbon 
atom pairs in the unit cell ($W = 42.6 \AA$) 
under a magnetic field of $B = 5660$ Teslas.
{\em Left panel:} 
In the upper figure we represent the edge state wavefunctions for the 
occupied and unoccupied $\uparrow$-spin and $\downarrow$-spin 
states when hybridization is not included. 
The lower figure represents the coefficients of the 
hybridized domain wall state.   All wavefunctions are 
plotted at the Bloch wavevector indicated by arrows 
on the right panel.
{\em Right panel:}
Quantum hall edge state bands 
with collinear spin states represented by  
black and red symbols and the non-collinear spin edge states
formed by optimizing the hybridization of up and down spin orbitals 
represented by the blue symbols.  We observe a clear
single-particle energy gap related to the energy gained 
by forming the non-collinear domain wall structure.
}
\end{figure}

\section{Summary and discussions}
\label{discussion}
We have carried out a mean-field study of graphene's $\nu=0$ ground state
which aims to shed light on the character of the interaction-induced
gap that appears experimentally.  The fact that the ground state 
has a charge gap at this filling factor can be established essentially rigorously\cite{yang} within the
continuum model often used to describe graphene.  When interaction induced mixing 
between $n=0$ and $|n| \ne 0$ Landau levels is neglected in the continuum model,
the family of broken symmetry states is related by 
arbitrary spin and valley pseudospin rotations and includes both fully spin-polarized 
$F$ and spin (SDW) and charge (CDW) wave states.  Our effort to determine the ground state 
character when Landau-level mixing and lattice effects are included, is motivated by
the observation of magnetic-field-induced insulating transport properties and by the expectation that 
transport properties in the quantum Hall regime should be very different for F, SDW, and CDW states.

The lattice model we study has two phenomenological parameters, a relative dielectric constant 
$\epsilon_r$ and an on-site interaction parameter $U$.  The most appropriate values of both 
parameters are somewhat uncertain.  Our two main findings are that 
i) Landau-level mixing effects favor the density-wave states over the ferromagnetic state and 
ii) The competition between CDW and SDW states is sensitive to the relative strength of 
on-site and inter-site electron-electron interactions and hence on the product $\epsilon_r \cdot U$.
Large values of this product increase the direct mean-field energy cost of CDW order and 
favor the SDW state.   Exchange energies are stronger for density wave states than for
ferromagnetic states because order within the $n=0$ level, induces order in the 
full negative $n$ Landau levels in the former case, but not in the latter case.
For graphene on SiO$_2$ and other typical substrates, sensible values for $\epsilon_r$ and 
$U$ suggest that the field-induced state at $n=0$ is a SDW state.  CDW states could occur 
in suspended graphene samples. 

The atomic value of the on-site interaction term in carbon is $U \sim 10 {\rm eV}$,
so graphene values should be smaller.  Most of the illustrative calculations we 
have described use either $U =4 eV$ or $U = 5 eV$.
The dielectric constant is $\epsilon_r \sim 1$ for free standing graphene.
For a SiO$_2$ substrate, with dielectric constant $\epsilon_r \sim 4.5$,
the effective 2D dielectric constant at a substrate/vacuum interface is $\epsilon_r \sim 2.5$.
Because the Hartree-Fock approximation overestimates the strength of exchange interactions, it
can be argued that somewhat larger values of $\epsilon_r$ are appropriate - perhaps $\sim 2$ 
for free standing graphene and $\epsilon_r \sim 4$ or $5$ for graphene on a substrate.
Because of these uncertainties we view $U$ and $\epsilon_r$ 
as effective parameters whose values are somewhat uncertain and 
have made energy comparisons over a wide range of values.  

We are able to complete our calculations only at magnetic field strengths 
much stronger than those available experimentally.  Partly to verify that the 
field-dependence is systematic, and 
partly in an effort to estimate the magnetic field strengths necessary for 
Zeeman energies to drive the system into a F state, we have fit the 
energy differences between F and density-wave states to the form
$\Delta E = C_{3/2} B^{3/2} + C_{2} B^2$.  Only the first term
can be present for a continuum model with Coulomb interactions, and we 
find that this term is indeed dominant.  
We find that Zeeman coupling at typical fields can change the nature of the state 
only in the parameter range where the crossover from CDW to SDW states occurs.  
Although we assume collinear states in analyzing the SDW/F competition, we presume
that the SDW to F crossover is actually a continuous one in which the 
antiparallel spins on opposite sublattices are smoothly rotated until they are aligned. 
Even if the bulk gap remains open the rotation of the spins will bring about a progressive 
closing of edge state band gaps, until the gap is 
completely closed in the F configuration. 
The modulation of the edge charge gap due to Pauli paramagnetism might be detected in 
edge transport experiments. 
Since a relatively strong magnetic field is required to induce the 
insulating state in typical samples, we do not expect that 
it will be easy to produce parallel fields large enough to turn 
the system metallic while maintaining this minimum perpendicular field. 
Nevertheless, a study of the parallel-field dependence of transport properties
is likely to hint at the nature of the underlying broken symmetry state of the system.

All of these results ignore the influence of disorder.  
Experiments appear to show that the transition to the insulating state occurs
at magnetic fields above some critical value that becomes smaller 
when the sample is cleaner.  This is expected, since disorder favors 
states without broken symmetries.  
The relationship between the minimum field and the mobility was carefully examined some time ago,\cite{nomura}
but can be crudely described using the following simple argument.
Assuming uncorrelated scatterers and using the Fermi golden rule the mobility is 
$\mu \propto 1/ V^2_{dis}$ where $V_{dis}$ is the typical energy scale of disorder.
The crossover occurs when disorder strength equals the interaction
energy scale $V_{dis} = U_{int} \propto B^{3/2}$,
therefore the critical magnetic field between samples with different
mobility are related through $B^{\prime}_{c} / B_{c}   = \left( \mu / \mu^{\prime} \right)^{1/3}$.
One physical picture of the strong disorder limit asserts that current flows along 
domain walls which separate disorder-induced electron-hole puddles\cite{sarma} 
 domain walls along which
current carrying states with $\nu \neq 0$ can dominate bulk transport
and suppress the divergent resistivity.
In graphene on substrates the mobility values range 
between 2000 and 25,000 $cm^{2} V^{-1} s^{-1}$, and values as high as
230,000 $cm^{2} V^{-1} s^{-1}$ have been achieved in annealed suspended samples \cite{bolotin}.
Typical crtical fields in samples on substrates are in the 20 to 30 Tesla range.
From the above argument we can expect that the critical magnetic fields in suspended samples should
be roughly 2 to 5 times lower.

An important goal of our work was to shed light on the character of the $\nu=0$ 
edge states.  We examined the electronic structure of armchair ribbons with CDW, SDW, and F states,
finding that edge states in the gap are absent for both CDW and SDW solutions.
Given that the field-induced insulating state appears to have 
an {\em extremely} large resistance once established, it appears likely to us that the 
experimental state does not have edge states and that it therefore must be 
a density wave, as suggested by these calculations.  If so, a study of 
the influence of magnetic field tilting might be able to distinquish between SDW and 
CDW states.

{\em Acknowledgments.}  
We acknowledge helpful discussions with R. Bistritzer, P. Cadden-Zimansky,
Y. Zhao, K. Bolotin, F. Ghahari, P. Kim
and financial support from Welch Foundation, NRI-SWAN, ARO, DOE
and the Spanish Ministry of Education through MEC-Fulbright program.
We thank the assistance and computer hours provided
by the Texas Advanced Computing Center (TACC).

\end{document}